\documentclass[12pt]{article}
\usepackage{graphicx}

\def \hf{\frac{1}{2}}

\def \bea{\begin{eqnarray}}
\def \beq{\begin{equation}}

\def \eea{\end{eqnarray}}
\def \eeq{\end{equation}}

\def \({\left(}
\def \){\right)}
\def \[{\left[}
\def \]{\right]}

\begin{document}
\rightline{TECHNION-PH-14-15}
%\rightline{October 2014}
%\rightline{arXiv:1410.xxxx}
%
\vskip 10mm
\centerline{\bf High-precision $D^0 \to V^+P^-$ -
more accurate than $D^0\to P^+P^-$}
\bigskip
\centerline{Michael Gronau}
\medskip
\centerline{\it Physics Department, Technion -- Israel Institute of Technology}
\centerline{\it Haifa 3200, Israel}
\bigskip
\begin{quote}
Recently we derived a nonlinear U-spin amplitude relation for $D^0\to P^+P^-$, 
$P=\pi, K$, predicted to hold up to fourth order U-spin breaking terms of order 
$10^{-3}$. Here we study a similar relation for $D^0\to V^+P^-, V =\rho, K^*(892), 
P = \pi, K$, expected to hold at an even higher accuracy of order $10^{-4}$. 
We confirm this prediction in spite of a large experimental 
error of about $20\%$ in the amplitude of $D^0\to K^{*+}\pi^-$.  We also 
comment briefly on U-spin breaking in $D^0\to P^+V^-$.
\end{quote}
\bigskip

U-spin symmetry has been known for a long time to play an important role in
obtaining approximate relations among hadronic $D$ meson decay amplitudes. 
In certain cases these relations were found to be violated by large symmetry breaking 
corrections. For instance, shortly after the discovery of charm in November 
1974, the following zeroth order U-spin relations were shown to hold among amplitudes 
for $D^0$ decays to pairs involving a charged pion or kaon~\cite{Kingsley:1975fe},
\bea\label{U}
& & A(D^0\to \pi^+K^-) : A(D^0\to K^+K^-) : A(D^0\to \pi^+\pi^-) : A(D^0 \to K^+\pi^-) 
\nonumber\\
&  & 
= \cos^2\theta_C : \cos\theta_C\sin\theta_C : -\cos\theta_C\sin\theta_C : -\sin^2\theta_C~,
\eea
where $\theta_C$ is the Cabibbo angle~\cite{Cabibbo:1963yz}, 
$\tan\theta_C = 0.23125 \pm 0.00082$~\cite{Agashe:2014kda}.

One of these relations, $|A(D^0\to K^+K^-)| = |A(D^0\to \pi^+\pi^-)|$ is broken 
experimentally by about $80\%$. A very early rudimental interpretation of this 
anomaly was suggested in the framework 
of flavor SU(3) breaking~\cite{Savage:1991wu}. Global studies within broken flavor SU(3)
of $D$ meson decays to two pseudoscalar mesons and to a pair of vector and pseudoscalar meson have been made in Ref.~\cite{Bhattacharya:2008ss}. 
A mechanism involving constructive interference of U-spin breaking penguin and 
tree (current-current) amplitudes has been suggested in 
Ref.~\cite{Bhattacharya:2012ah} to explain this anomaly in the framework of 
first order U-spin breaking. 
%MG separate Ref.
See also~\cite{Feldmann:2012js}.
 
 Very recently we have applied perturbatively high order U-spin breaking  to the relations 
 (\ref{U}), expanding amplitudes up to fourth order in powers of two distinct U-spin breaking 
 parameters, ${\rm Re}\epsilon_1=0.05, {\rm Re \epsilon_2}=0.30$~\cite{Gronau:2013xba}. 
 Confronting predictions for ratios of amplitudes by their measurements we 
 confirmed that second and fourth order U-spin breaking terms obey the 
 required hierarchy with respect to these first order parameters. Namely, second order 
 U-spin breaking corrections are a few times $10^{-2}$ while fourth order terms are of order 
 $10^{-3}$. 
 
 One particularly interesting high-precision relation was predicted in \cite{Gronau:2013xba}
 among four ratios of amplitudes each of which equals one in the U-spin symmetry limit. 
 Denoting
 \beq
A_1\equiv |A(\pi^+K^-)|\,,~~~~A_2\equiv |A(K^+\pi^-)|\,,~~~~A_3\equiv|A(\pi^+\pi^-)|\,,~~~
A_4\equiv |A(K^+K^-)|~,
 \eeq 
we defined four ratios,
\beq\label{Ri}
R_1\equiv\frac{A_2}{A_1\tan^2\theta_C}\,,~~~~R_2 \equiv \frac{A_4}{A_3}\,,~~~
R_3\equiv\frac{A_3 + A_4}{A_1\tan\theta_C + A_2\tan^{-1}\theta_C}\,,~~~
R_4\equiv\sqrt{\frac{A_3A_4}{A_1A_2}}\,.
\eeq
These ratios were shown to obey the following  nonlinear relation up to fourth order 
U-spin breaking,
\beq\label{NLrel}
\Delta R \equiv R_3 - R_4 + \frac{1}{8}\left[(\sqrt{2R_1-1} - 1)^2 - (\sqrt{2R_2 - 1} -1)^2\right]=
{\cal O}(\epsilon^4)~.
\eeq
The parameter $\epsilon$ represents generically the above-mentioned U-spin 
breaking parameters, $\epsilon_1$ and $\epsilon_2$.
Considering also first order isospin breaking, we have noted in~\cite{Gronau:2013xba}
that the right hand side may also include a tiny term suppressed by both isospin breaking and 
U-spin breaking parameters. 

The proof of (\ref{NLrel}) for the four amplitudes of $D^0 \to P^+P^- (P=\pi, K)$ was based 
on the U-spin vector behavior of the weak Hamiltonian, and on the
fact that $D^0$ is a U-spin singlet while each of the two pairs $(\pi^-, K^-)$ and $(K^+, -\pi^+)$ is 
a  U-spin doublet. 
Replacing the second pair by a U-spin doublet pair of vector mesons $(K^{*+}, -\rho^+)$, one 
may immediately apply this relation to other four decay amplitudes for $D^0 \to V^+P^- 
(V= \rho, K^*, P=\pi, K)$. Thus denoting 
\beq
A'_1\equiv |A(\rho^+K^-)|\,,~~~~A'_2\equiv |A(K^{*+}\pi^-)|\,,~~~~A'_3\equiv|A(\rho^+\pi^-)|\,,~~~
A'_4\equiv |A(K^{*+}K^-)|~,
\eeq 
and defining
\beq\label{Ri'}
R'_1\equiv\frac{A'_2}{A'_1\tan^2\theta_C}\,,~~~~R'_2 \equiv \frac{A'_4}{A'_3}\,,~~~
R'_3\equiv\frac{A'_3 + A'_4}{A'_1\tan\theta_C + A'_2\tan^{-1}\theta_C}\,,~~~
R'_4\equiv\sqrt{\frac{A'_3A'_4}{A'_1A'_2}}\,,
\eeq
one has
\beq\label{NLrel'}
\Delta R' \equiv R'_3 - R'_4 + \frac{1}{8}\left[(\sqrt{2R'_1-1} - 1)^2 - (\sqrt{2R'_2 - 1} -1)^2\right]=
{\cal O}(\epsilon'^4)~.
\eeq
Here $\epsilon'$ represents generically two U-spin breaking parameters, $\epsilon'_1$ and 
$\epsilon'_2$,  defined for $D^0 \to V^+P^-$ in analogy to $\epsilon_1$ and $\epsilon_2$ 
defined for $D^0 \to P^+P^-$.

% This is Table I
\begin{table}[t]
\caption{Branching fractions and amplitudes for $D^0\to P^+P^-$ 
decays~\cite{Agashe:2014kda}\label{tab:PP}} 
\begin{center}
\begin{tabular}{c c c c c} \hline \hline
Decay mode  & Branching fraction & $p$ (${\rm GeV}/c$) & 
$A=\sqrt{{\cal B}/p}\,({\rm GeV}/c)^{-1/2}$ & \\ \hline
 $D^0\to \pi^+K^-$  &  ${\cal B}_{\pi\hskip-0.5mmK}=
 (3.88\pm 0.05)\hskip-1mm\times\hskip-1mm10^{-2}$  & $0.861$ & 
 $1.078{\cal B}_{\pi\hskip-0.5mmK}^{1/2}$ & $A_1$\\
 $D^0 \to K^+\pi^- $ &  $(3.56 \pm 0.06)\hskip-1mm\times\hskip-1mm10^{-3}
 {\cal B}_{\pi\hskip-0.5mmK}$ & $0.861$ & $(0.06430\pm 0.00054)
 {\cal B}_{\pi\hskip-0.5mmK}^{1/2}$ & $A_2$ \\ 
 $D^0 \to \pi^+\pi^-$ & $(3.59\pm 0.06)\hskip-1mm\times\hskip-1mm10^{-2}
 {\cal B}_{\pi\hskip-0.5mmK}$ & $0.922$ & 
 $(0.1973\pm 0.0016){\cal B}_{\pi\hskip-0.5mmK}^{1/2}$ & $A_3$ \\ 
 $D^0 \to K^+K^-$ & $(10.10 \pm 0.16)\hskip-1mm\times\hskip-1mm10^{-2}
 {\cal B}_{\pi\hskip-0.5mmK}$ & $0.791$ & 
 $(0.3573\pm 0.0028){\cal B}_{\pi\hskip-0.5mmK}^{1/2}$ & $A_4$ \\
 \hline\hline
 \end{tabular}
\end{center}
\end{table}
%
 % This is Table II
\begin{table}[h]
\caption{Branching fractions and amplitudes for $D^0\to V^+P^-$ 
decays~\cite{Agashe:2014kda}\label{tab:VP}} 
\begin{center}
\begin{tabular}{c c c c c} \hline \hline
Decay mode  & Branching fraction & $p$ (${\rm GeV}/c$) & 
$A'=\sqrt{{\cal B}/p^3}\,({\rm GeV}/c)^{-3/2}$ & \\ \hline
 $D^0 \to \rho^+K^-$ & $0.108 \pm 0.007$ & $0.675$ & $0.593 \pm 0.019$& $A'_1$\\
 $D^0 \to K^{*+}\pi^-$ & $(3.42^{+1.80}_{-1.02})\hskip-1mm\times\hskip-1mm 
 10^{-4}$ & $0.711$ & $0.0308^{+0.0081}_{-0.0046}$ & $A'_2$ \\
 $D^0 \to \rho^+\pi^-$ & $(9.8 \pm 0.4)\hskip-1mm\times\hskip-1mm 10^{-3}$ 
 & $0.764$ & $0.148 \pm 0.003$ & $A'_3$ \\
 $D^0 \to K^{*+}K^-$ & $(4.38 \pm 0.21)\hskip-1mm\times\hskip-1mm 10^{-3}$ 
 & $0.610$ & $0.1389 \pm  0.0033$ & $A'_4$ \\
 \hline \hline
\end{tabular}
\end{center}
\end{table}

We will now compare the current experimental status of the two predicted amplitude 
relations~(\ref{NLrel}) and (\ref{NLrel'}), paying special attention to experimental errors. 
Hadronic decay amplitudes are obtained from measured branching ratios ${\cal B}$ by 
eliminating U-spin breaking phase space factors. For s-wave $D\to P^+ P^-$ decays 
and for  p-wave $D^0 \to V^+P^-$ decays one has
\beq
|A| = M_D\sqrt{\frac{8\pi{\cal B}}{\tau_D\,p}}~,~~~~~
|A'| = M_D\sqrt{\frac{8\pi{\cal B}}{\tau_D\,p^3}}~.
\eeq  
Here $p$ is the center-of-mass 3-momentum of each final particle, while $M_D$ and 
$\tau_D$ are the $D$ meson mass and its lifetime. Since we are only concerned with 
ratios of amplitudes we will disregard a common factor $M_D\sqrt{8\pi/\tau_D}$ which 
cancels in these ratios. 

Values for measured branching ratios, center-of-mass 
momenta, and amplitudes defined in the above manner are quoted in Tables \ref{tab:PP} 
and \ref{tab:VP} for $D^0\to P^+P^-$ and $D^0 \to V^+P^-$, respectively. 
Focusing at this point on experimental errors, we note that 
we have included no error in the amplitude for the Cabibbo-favored (CF) decay $D^0\to \pi^+K^-$.
All other three $D^0\to P^+P^-$ branching ratios including errors have been measured relative 
to this process~\cite{Agashe:2014kda}. The three errors in amplitudes for singly Cabibbo-suppressed (SCS) decays $D^0 \to \pi^+\pi^-, D^0\to K^+K^-$ and for the doubly Cabibbo-suppressed (DCS) decay $D^0\to K^+\pi^-$ are all around $0.8\%$, below the level 
of one percent. The high precision achieved recently in measuring the DCS amplitude is remarkable. It required time-dependent 
separation between this highly suppressed decay and $D^0$-$\bar D^0$ mixing followed 
by the CF decay $\bar D^0 \to K^+\pi^-$~\cite{Aaltonen:2013pja}. 

Considering current errors in $D^0\to V^+P^-$ amplitudes we note that the relative 
errors in CF and SCS amplitudes are reasonably small, between two 
and three percent. On the other hand, the relative error in the DCS amplitude 
$A(D^0\to K^{*+}\pi^-)$, obtained through a Dalitz plot analysis of 
$D^0\to K_S\pi^+\pi^-$~\cite{Asner:2003uz}, is quite large - $^{+26\%}_{-15\%}$. 
This large asymmetric error limits considerably the precision of 
$R'_1, R'_3$ and  $R'_4$ occurring in Eq.~(\ref{NLrel'}). This would seem to prohibit a precise test 
of this high order U-spin relation using current data. We will show that this is actually 
not exactly the case. That is, in spite of a dominant large error in $\Delta R'$ originating in
$A(D^0\to K^{*+}\pi^-)$, the amplitude relation (\ref{NLrel'}), in which one neglects fourth order 
U-spin breaking, holds even better than the corresponding relation in $D^0 \to P^+P^-$. 
Using current experimental amplitudes we will now calculate the quantities $\Delta R$ and 
$\Delta R'$ defined in (\ref{NLrel}) and (\ref{NLrel'}) including their errors. 

Taking values of amplitudes given in Tables \ref{tab:PP} and \ref{tab:VP} we first calculate,
separately for $D^0\to P^+P^-$ and $D^0\to V^+P^-$,  
the four ratios of amplitudes defined in Eqs.~({\ref{Ri}) and (\ref{Ri'}).
%MG added following referee's  comment
The measured branching ratios, obtained in independent analyses for different 
two-body $P^+P^-$ and three-body $V^+P^-$ final states, involve no error correlations.
Therefore we compute errors in ratios by adding 
in quadrature errors due to the relevant amplitudes. We find
\beq\label{Rinum}
R_1 = 1.115 \pm 0.012\,,~~~~R_2 = 1.811 \pm 0.020\,,~~~
R_3 = 1.052 \pm 0.008\,,~~~~R_4 = 1.008 \pm 0.007\,,
\eeq
and
\beq\label{R'inum}
R'_1 = 0.971^{+0.257}_{-0.148}\,,~~~~~R'_2 = 0.939 \pm 0.029\,,~~~~
R'_3 = 1.061^{+0.082}_{-0.140}\,,~~~~R'_4 = 1.061^{+0.083}_{-0.142}\,.
\eeq
Errors in $\Delta R$ and $\Delta R'$ caused by errors in amplitudes are also added 
in quadrature [rather than using the errors in (\ref{Rinum}) and (\ref{R'inum})]:
\beq
\Delta R = (-3.2 \pm 0.4)\times 10^{-3}~,~~~~
\Delta R' = (0.2^{+3.2}_{-5.5})\times 10^{-4}~.
\eeq
Thus, in spite of its current huge relative error, $\Delta R'$ is significantly smaller than 
$\Delta R$. We have confirmed that the dominant uncertainty in $\Delta R'$ originates in
the large experimental error in $|A(D^0 \to K^{*+}\pi^-)|$. Neglecting this error we find 
$\Delta R' = (0.2 \pm 0.4)\times 10^{-4}$, which is smaller than $\Delta R$ by more than 
an order of magnitude. Can one explain this somewhat unanticipated situation?

As is explicit on the right hand sides of (\ref{NLrel}) and (\ref{NLrel'}), $\Delta R$ and 
$\Delta R'$ are expected 
to be tiny but nonzero due to fourth order U-spin breaking terms and possible 
corrections which break both isospin and U-spin. It is remarkable but not unexpected,
we will now show, that these symmetry breaking terms are smaller in $D^0\to V^+P^-$ 
than in $D^0 \to P^+P^-$.

Using relations derived in Ref.~\cite{Gronau:2013xba} between two U-spin breaking parameters,
$\epsilon^{(,)}_{1,2}$ and the two ratios $R^{(,)}_{1,2}$, we calculate for 
$D^0\to P^+P^-$ and $D^0\to V^+P^-$:
\beq\label{Reep}
{\rm Re}\,\epsilon_1=\hf\left(\sqrt{2R_1 - 1} -1\right)=0.054 \pm 0.005\,,~~
{\rm Re}\,\epsilon_2=\hf\left(\sqrt{2R_2 - 1} -1\right) = 0.310  \pm 0.006\,,
\eeq
\beq\label{Reep'}
{\rm Re}\,\epsilon'_1=\hf\left(\sqrt{2R'_1 - 1} -1\right) = -0.015^{+0.118}_{-0.083}\,,~~
{\rm Re}\,\epsilon'_2=\hf\left(\sqrt{2R'_2 - 1} -1\right) =  -0.032 \pm 0.016\,.
\eeq

While the U-spin breaking parameter $\epsilon_2$ in $D^0 \to P^+P^-$ is quite
large, ${\rm Re}\,\epsilon_2 \simeq 0.30$, the corresponding parameter in $D^0 \to V^+P^-$ is an
order of magnitude smaller, $-{\rm Re}\,\epsilon'_2 \sim 0.01 - 0.05$.
Namely, a U-spin breaking mechanism involving constructive interference of penguin and tree amplitudes, suggested in Ref.~\cite{Bhattacharya:2012ah} to account for the large 
value of $R_2 \equiv |A(D^0 \to K^+K^-)|/|A(D^0\to \pi^+\pi^-)|$, is not at work in 
$R'_2 \equiv |A(D^0 \to K^{*+}K^-)|/|A(D^0 \to \rho^+\pi^-)|$. 
The ratio $R'_1$ does not involve a penguin contribution and is likely to deviate from one by 
no more than about $20\%$, typical for U-spin breaking in tree amplitudes.  (This may be demonstrated, for instance, by a simple model calculation based on factorization,
using as input the ratio of $K^*$ to $\rho$ decay constants and the ratio of $D\to \pi$ to
$D \to K$ form factors for $q^2=m^2_{K^*}$ and $q^2=m^2_\rho$, respectively.) Consequently, one
expects $|{\rm Re}\,\epsilon'_1| \le 0.1$, in agreement with (\ref{Reep'}). This parameter could be smaller, around $0.05$, just like ${\rm Re}\,\epsilon_1$. Taking these small values of
${\rm Re}\,\epsilon'_1$ and ${\rm Re}\,\epsilon'_2$ as typical first order U-spin breaking in 
$D^0\to V^+P^-$ amplitudes implies that a fourth order symmetry breaking term in 
$\Delta R'$ is at most $10^{-4}$. An isospin breaking  term suppressed also by
U-spin is expected to be a few times $10^{-4}$.

Also, the ratios $R'_3$ and $R'_4$ which involve large experimental errors are predicted 
to equal one up to second order U-spin breaking corrections~\cite{Gronau:2013xba}.
In view of the above small values of ${\rm Re}\,\epsilon'_1$ and ${\rm Re}\,\epsilon'_2$, we
expect these corrections to be at most a few percent. 

High-precision relations of the form (\ref{NLrel}) and (\ref{NLrel'}) hold whenever final 
states in $D^0$ decays involve pairs of U-spin doublet mesons. We do not study the 
relation in decays to
$P^+V^-$ because no data are available at this time for the DCS decay $D^0 \to K^+\rho^-$.
We checked that the magnitude of a corresponding U-spin breaking parameter given by 
a ratio of two SCS $D^0 \to P^+V^-$ amplitudes is intermediary between corresponding 
parameters in $D^0\to P^+P^-$ and $D^0 \to V^+P^-$. Using branching ratios given 
in Table \ref{tab:PV}, we calculate
\beq
R''_2 \equiv \frac{|A(D^0 \to K^+K^{*-})|}{|A(D^0 \to \pi^+\rho^-)|} = 0.786 \pm 0.036
~~~{\rm implying}~~~~{\rm Re}\,\epsilon''_2 = -0.122 \pm 0.024~.
\eeq
This value of ${\rm Re}\,\epsilon''_2$ should be compared with 
${\rm Re}\,\epsilon_2=0.310\pm 0.006$ and ${\rm Re}\,\epsilon'_2=-0.032\pm0.016$
in $D^0\to P^+P^-$ and $D^0\to V^+P^-$, respectively.
%
 % This is Table III
\begin{table}[h]
\caption{Branching fractions and amplitudes for SCS $D^0\to P^+V^-$ 
decays~\cite{Agashe:2014kda}\label{tab:PV}} 
\begin{center}
\begin{tabular}{c c c c } \hline \hline
Decay mode  & Branching fraction & $p$ (${\rm GeV}/c$) & 
$A''=\sqrt{{\cal B}/p^3}\,({\rm GeV}/c)^{-3/2}$ \\ \hline
 $D^0 \to \pi^+\rho^-$ & $(4.96 \pm 0.24)\hskip-1mm\times\hskip-1mm 10^{-3}$ 
 & $0.764$ & $0.1055 \pm 0.0026$  \\
 $D^0 \to K^+ K^{*-}$ & $(1.56 \pm 0.12)\hskip-1mm\times\hskip-1mm 10^{-3}$ 
 & $0.610$ & $0.0829 \pm  0.0032$ \\
 \hline \hline
\end{tabular}
\end{center}
\end{table}

In conclusion, we have studied the current experimental status of a nonlinear precision 
relation predicted for $D^0\to V^+P^-$ decay amplitudes, comparing its precision to that of a 
similar relation in $D^0 \to P^+P^-$ decays. We found that while the
latter relation is violated at a very low level of $10^{-3}$, the relation for $D^0 \to V^+ P^-$ 
holds at an even higher precision.  We have shown that the correction to this relation,
 representing a fourth order U-spin breaking term or an isospin breaking term suppressed 
 also by U-spin, is consistent with first order U-spin 
 breaking parameters that are smaller in $D^0 \to V^+P^-$ than in $D^0\to P^+P^-$. 
 
 A small uncertainty in the $D^0 \to V^+P^-$ relation, at a level less than $10^{-3}$, is 
 due to the current large experimental error in ${\cal B}(D^0 \to K^{*+}\pi^-)$. This error is 
expected to be reduced in future experiments by the LHCb and Belle II 
collaborations. It will be interesting to watch future improvements in this measurement,
providing more precise determinations of 
%MG added after submission to PRD
${\rm Re}\,\epsilon'_1$, of second order U-spin breaking terms  in $R'_3, 
R'_4$ at a percent level, and of a fourth order term in $\Delta R'$ at a level of $10^{-4}$.
The high-precision relations discussed in this report within the Standard Model provide 
useful constraints on $|\Delta C|=1$ new physics operators~\cite{Gronau:2014pra}.

I wish to thank David Asner for drawing my attention to the two papers in 
Ref.~\cite{Asner:2003uz} and Shlomo Dado for checking error calculations.

\end{document}